\documentclass[10pt,twocolumn]{article}
\usepackage{graphicx} 
\graphicspath{{figures}} 
\usepackage[font=small,labelfont=bf]{caption} 

\usepackage{xcolor}
\definecolor{black}{HTML}{212427}
\definecolor{blue}{HTML}{0563C1}
\definecolor{brightred}{HTML}{FF0000} 

\usepackage{hyperref,nameref}
\hypersetup{colorlinks=true,allcolors=blue}
\newcommand{\rref}[2]{\hyperref[#1]{\ref{#1}#2}} 

\color{black}
\usepackage[margin=0.67in]{geometry} 

\usepackage[switch]{lineno}   
\usepackage{amsmath,amssymb}  
\usepackage{etoolbox} 
\newcommand*\linenomathpatch[1]{%
  \cspreto{#1}{\linenomath}%
  \cspreto{#1*}{\linenomath}%
  \csappto{end#1}{\endlinenomath}%
  \csappto{end#1*}{\endlinenomath}%
}
\linenomathpatch{equation}
\linenomathpatch{gather}
\linenomathpatch{multline}
\linenomathpatch{align}
\linenomathpatch{alignat}
\linenomathpatch{flalign}

\usepackage{fancyhdr,lastpage} 
\usepackage{titlesec}
\titleformat{\section}{\Large\bfseries}{}{0mm}{}
\titleformat{\subsection}{\bfseries}{}{0mm}{}
\titlespacing{\section}{0pt}{\baselineskip}{0pt}
\titlespacing*{\section}{0pt}{\baselineskip}{0pt}
\titlespacing*{\subsection}{0pt}{\baselineskip}{0pt}

\renewcommand{\v}[1]{\boldsymbol{\mathbf{#1}}} 
\newcommand{\uv}[1]{\boldsymbol{\mathbf{\hat{#1}}}} 
\renewcommand{\l}[0]{\left} 
\renewcommand{\r}[0]{\right} 
\let\f=\frac 
\renewcommand{\t}[1]{\text{#1}} 

\newcommand{\rev}[1]{{#1}} 

\usepackage[
  backend=biber,        
  autocite=superscript, 
  sorting=none,         
  style=nature,   
  maxnames=6,         
  minnames=1,         
  giveninits=false,     
  autopunct=false,      
  hyperref=true         
]{biblatex}

\DeclareFieldFormat{labelnumberwidth}{\;[#1]\!\!\!\!} 
\renewbibmacro{in:}{} 
\AtNextBibliography{\small} 
\AtEveryBibitem{\clearfield{month}} 

\usepackage[shortcuts, abbreviations,]{glossaries-extra}
\newabbreviation{SRO}{SRO}{short-range order}
\newabbreviation{SS}{SS}{solid solution}
\newabbreviation{fcc}{fcc}{face-centered cubic}
\newabbreviation{bcc}{bcc}{body-centered cubic}
\newabbreviation{hcp}{hcp}{hexagonal close-packed}
\newabbreviation{WC}{WC}{Warren-Cowley}
\newabbreviation{1CP}{1CP}{first coordination polyhedron}
\newabbreviation{KL}{KL}{Kullback-Leibler}
\newabbreviation{ML}{ML}{machine learning}
\newabbreviation{MC}{MC}{Monte Carlo}
\newabbreviation{1NN}{1NN}{first nearest neighbors}

\glsdisablehyper 

\usepackage[]{cleveref} 
\creflabelformat{equation}{#2\textup{#1}#3} 

\usepackage{tabularx}
\usepackage{booktabs}
\newcolumntype{Y}{>{\raggedleft\arraybackslash}X} 

\begin{document}

\twocolumn[
  \begin{center}
    \large
     \textbf{Chemical-motif characterization of short-range order with E(3)-equivariant graph neural networks}
  \end{center}
  Killian Sheriff$^1$,
  Yifan Cao$^1$, and
  Rodrigo Freitas$^1${\footnotemark[1]} \\
  $^1$\textit{\small Department of Materials Science and Engineering, Massachusetts Institute of Technology, Cambridge, MA, USA} \\

  {\small Dated: \today}

  \vspace{-0.15cm}
  \begin{center}
    \textbf{Abstract}
  \end{center}
  \vspace{-0.35cm}
  Crystalline materials have atomic-scale fluctuations in their chemical composition that modulate various mesoscale properties. Establishing chemistry--microstructure relationships in such materials requires proper characterization of these chemical fluctuations. Yet, current characterization approaches (e.g., Warren-Cowley parameters) make only partial use of the complete chemical and structural information contained in local chemical motifs. Here we introduce a framework based on E(3)-equivariant graph neural networks that is capable of completely identifying chemical motifs in arbitrary crystalline structures with any number of chemical elements. This approach naturally leads to a proper information-theoretic measure for quantifying chemical short-range order (SRO) in chemically complex materials, and a reduced representation of the chemical \rev{motif} space. Our framework enables the correlation of any per-atom property with their corresponding local chemical motif, thereby enabling the exploration of structure--property relationships in chemically-complex materials. Using the MoTaNbTi high-entropy alloy as a test system, we demonstrate the versatility of this approach by evaluating the lattice strain associated with each chemical motif, and computing the temperature dependence of chemical-fluctuations length scale.
  \vspace{0.4cm}
]
{
  \footnotetext[1]{Corresponding author (\texttt{rodrigof@mit.edu}).}
}

\begin{figure*}[!bt] 
  \centering
 \includegraphics[width=\textwidth]{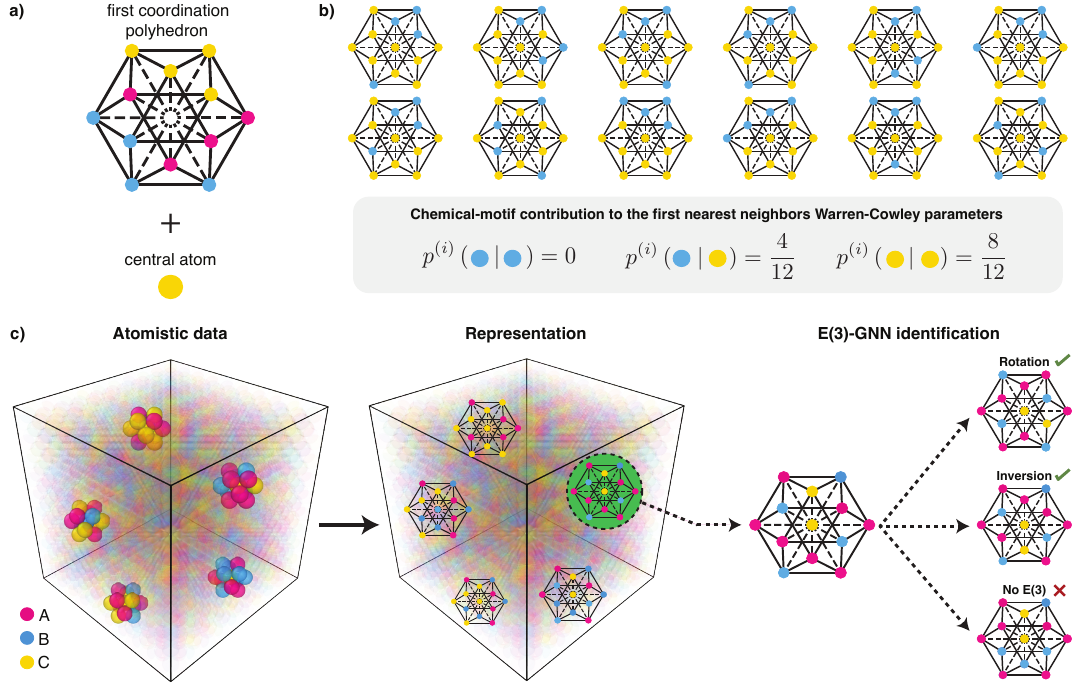}
  \caption{\label{fig:fig_framework}\textbf{Representation and identification of local chemical motifs.} \textbf{a)} A local chemical motif is defined by the central atom and its first coordination polyhedron. \textbf{b)} Distinct chemical motifs with the same average chemical composition contribute equally to \rev{first nearest neighbors} Warren-Cowley parameters. \textbf{c)} Illustration of the chemical-motif identification framework. Each atom in the system is first given awareness of its chemical environment by being represented as a local chemical motif. During the identification step, the graph representation of the motif is employed by an E(3)-equivariant graph neural network --- E(3)-GNN --- to identify equivalent motifs, i.e., motifs that can be transformed into each other by euclidean symmetries.}
\end{figure*}

\noindent Metallic\autocite{george_high-entropy_2019} and ceramic\autocite{oses_high-entropy_2020} alloys are often synthesized in phases where the chemical elements are dispersed almost randomly on a crystal lattice, namely \gls{SS} phases. Characterizing the spatial distribution of chemical elements at the atomic scale is critical in establishing structure--property relationships in \glspl{SS}. For example, spatial variations in chemistry cause strengthening via solute--dislocation like interactions\autocite{varvenne_theory_2016}, while the percolation of locally passive chemical regions is associated with corrosion resistance\autocite{xie_percolation_2021}. In a truly random \gls{SS} the spatial distribution of chemical elements is defined by the underlying geometry of the crystal lattice and its associated symmetries. However, in real materials, thermal effects induce a trade-off between enthalpy and entropy that favors low-energy chemical motifs\autocite{yang_coordination_2008, polak_local_2003, clapp_atomic_1971,sheriff2023quantifying} (fig.~\rref{fig:fig_framework}{a}), thereby affecting the spatial distribution of chemical elements. This deviation from randomness is known as chemical \gls{SRO}.

The state of \gls{SRO} is often characterized using the \gls{1NN} \gls{WC} parameters\autocite{cowley_approximate_1950, ding_tunable_2018, yu_theory_2022, sun_effect_2022, li_strengthening_2019, ghosh_short-range_2022, walsh_magnetically_2021,feng_effects_2017, du_chemical_2022, yin_atomistic_2021,kostiuchenko_impact_2019,ghosh_short-range_2022}, defined as
\begin{equation}\label{eq:wc}
    \alpha_{\t{AB}} = 1-\frac{p(\t{A}|\t{B})}{c_{\t{A}}} =  1 - \frac{1}{c_{\t{A}}} \l[  \frac{1}{Nc_{\t{B}}} \times \sum_{i=1}^N p^{(i)}(\t{A}|\t{B}) \r],
 \end{equation}
where A and B are chemical elements, $c_\t{A}$ is the average concentration of atoms of type A, $N$ is the total number of atoms, and $p(\t{A}|\t{B})$ is the probability of finding an atom of type A in the \gls{1NN} shell of a B atom, which can be broken down into a per-chemical-motif contribution $p^{(i)}(\t{A}|\t{B})$. \rev{When considering only the 1NN, these $n_\t{c}(n_\t{c}+1)/2$ independent WC parameters (where $n_\t{c}$ is the number of chemical elements in the system)} provide an incomplete\autocite{sheriff2023quantifying} description of SRO \rev{at the first-neighbor shell} since distinct chemical motifs with the same chemical concentration are indistinguishable and contribute equally to the WC parameters. For example,  fig.~\rref{fig:fig_framework}{b} illustrates several chemical motifs with different bonding environments that all have the same contribution to WC parameters. Consequently, as shown in our previous work (ref.~\cite{sheriff2023quantifying}), reverse-engineered atomic configurations from \gls{WC} parameters yield unphysical degenerate solutions, and the lack of non-degenerate descriptors prevents the connection of per-atom properties (e.g., generalized stacking fault energy \autocite{utt_origin_2022} and magnetic moments\autocite{walsh_magnetically_2021}) with their corresponding chemical motif. \rev{Meanwhile, multi-shell WC parameters can be seen as a full characterization of the \textit{chemical concentration} variation with distance\autocite{MB_WC,Singh2015}. Despite the completeness of this description (for chemical concentration), understanding and quantifying the compatibility conditions among WC parameters across different shells remains a longstanding challenge\autocite{walsh_reconsidering_2023}.
}

To move beyond the characterization of bonding preferences provided by WC parameters, experimental efforts have employed transmission electron microscopy techniques\autocite{zhou_atomic-scale_2022, zhang_short-range_2020, chen_direct_2021, xu_determination_2023, hsiao_data-driven_2022, wang_chemical_2022,xu_correlating_2023,rakita_mapping_2023} to assess spatial correlations among atomic columns' chemistry. Yet, these methods are unable to access the complete 3D spatial chemical distribution, and the signals associated with \gls{SRO} may have originated from other atomistic features\autocite{walsh_extra_2023, joress_why_2023, coury_origin_2023}. Meanwhile, atom probe tomography approaches \autocite{hu_application_2022, moniri_three-dimensional_2023, li_machine_2023, li_convolutional_2021} are nascent and provide complete 3D characterization, but are still limited in accuracy and by anisotropic resolution\autocite{gault_spatial_2010, gault_atom_2021}. As experimental efforts evolve in their ability to capture the spatial distribution of chemical elements and correlate them with materials properties, they would benefit from a framework for the non-degenerate identification of chemical-motifs and quantification of SRO beyond WC parameters.

Here, we propose an approach to characterize the state of \gls{SRO} using all of the 3D atomistic information available. By employing \gls{ML} and group theory together, we are able to extend the local chemical motif method introduced in ref.~\cite{sheriff2023quantifying} into a framework that is applicable to arbitrary crystal lattices with any number of elements; the mathematical foundation for such generalization is provided.  This approach naturally leads to a proper information-theoretic measure for quantifying SRO, and a reduced --- but complete (i.e. non-degenerate) --- representation of the chemical \rev{motif} space. We demonstrate the application of this approach by identifying all possible chemical motifs in \gls{fcc}, \gls{bcc}, and \gls{hcp} systems containing up to five chemical elements, and quantitatively characterizing the state of \gls{SRO} in the \gls{bcc} MoTaNbTi high-entropy alloy using a machine learning potential.

\section{Results}

\subsection{Representation and enumeration of chemical motifs}

\begin{figure*}[!bt]
  \centering
 \includegraphics[width=\textwidth]{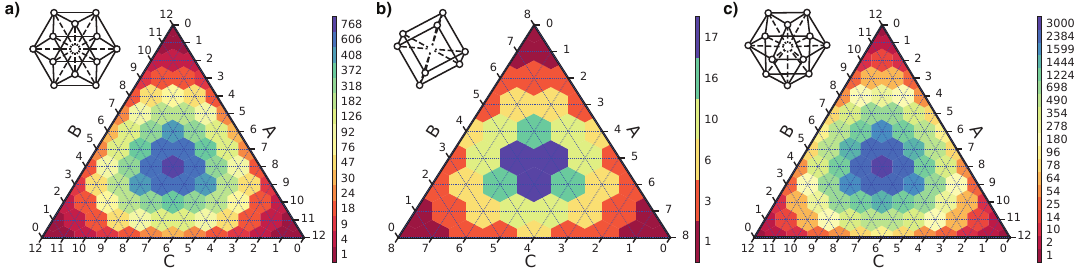}
  \caption{\label{fig:fig_pattern_inventory} \textbf{Counting of distinct chemical motifs in three-element crystal lattices}. The ternary diagrams indicate the chemical composition of the first coordination polyhedron (1CP, illustrated in fig.~\rref{fig:fig_framework}{a}) and the color bar on the right shows the number of distinct 1CPs for that corresponding composition. The number of distinct chemical motifs is obtained after multiplying by $n_\t{c} =3$ to account for the central atom type. This analysis was performed for the \textbf{a)} face-centered cubic, \textbf{b)} body-centered cubic, and \textbf{c)} hexagonal close-packed lattices.}
\end{figure*}

The local chemical motif $\mathcal{M}_i$ is defined as the group of atoms composed of a central atom $i$ and its \gls{1CP}, as illustrated in fig.~\rref{fig:fig_framework}{a}. The resulting polyhedron is the cornerstone of our atomic-scale analysis because it completely characterizes the local atomic environment surrounding atom $i$. In a \gls{fcc} lattice the \gls{1CP} (fig.~\rref{fig:fig_framework}{a}) takes the form of a cuboctahedron with octahedral symmetry point group $O_h$. This cuboctahedron is comprised of eight triangular faces, six square faces, and 12 vertices, representing the $N_\t{a}=12$ \gls{1NN} surrounding the central atom. Consequently, the chemical motif consists of 36 edges connecting 13 atoms (including the central atom). Out of these edges, 24 correspond to connections among the \gls{1NN}, while the remaining 12 edges account for the connections between the central atom and the \gls{1NN}. A similar description of the geometry of \gls{bcc} (fig.~\rref{fig:fig_pattern_inventory}{b}) and \gls{hcp} (fig.~\rref{fig:fig_pattern_inventory}{c}) motifs is provided in Supplementary Section 1.

A chemical motif $\mathcal{M}_i$ is deemed equivalent to another motif $\mathcal{M}_j$ (i.e., $\mathcal{M}_i \sim \mathcal{M}_j$) if they are related through euclidean symmetry operations. Conversely, two motifs are said to be distinct (i.e., $\mathcal{M}_i \nsim \mathcal{M}_j$) if they cannot be related to each other by any euclidean symmetry. Figure \rref{fig:fig_framework}{c} illustrates how a particular motif is equivalent to two other motifs and distinct from a third motif.

\begin{table*}[!bt]
 \begin{tabularx}{\linewidth}{l*{5}{Y}}
   \toprule
   \midrule
   \textbf{Number of chemical elements $(n_\t{c})$ \hspace{-0.3cm} } & 2 & 3 & 4 & 5 \\
   \bottomrule
 \end{tabularx}
 \begin{tabularx}{\linewidth}{l*{5}{Y}}
   \toprule
   \multicolumn{5}{l}{\textbf{Face-centered cubic (fcc)}}                                                 \\
   \midrule
   Number of possible motifs ($n_\t{c}^{13}$)                            & $8,192$ & $1,594,323$ & $67,108,864$ & $1,220,703,125$ \\
   Number of unique compositions $\l[\frac{n_\t{c}(n_\t{c}+11)!}{12!(n_\t{c}-1)!}\r]$ & $26$    & $273$      & $1,820$        & $9,100$       \\
   Number of distinct motifs $(N_\t{dm})$  & $288$   & $36,333$  & $1,432,480$    & $25,658,250$   \\
   \bottomrule
 \end{tabularx}
 \begin{tabularx}{\linewidth}{l*{5}{Y}}
   \toprule
   \multicolumn{5}{l}{\textbf{Body-centered cubic (bcc)}}                                    \\
   \midrule
   Number of possible motifs ($n_\t{c}^{9}$)                            & $512$ & $19,683$ & $262,144$ & $1,953,125$ \\
   Number of unique compositions $\l[\frac{n_\t{c}(n_\t{c}+7)!}{8!(n_\t{c}-1)!}\r]$  & $18$   & $135$    & $660$    & $2,475$     \\
   Number of distinct motifs $(N_\t{dm})$ & $44$  & $801$   & $7,984$  & $51,875$  \\
   \bottomrule
 \end{tabularx}
 \begin{tabularx}{\linewidth}{l*{5}{Y}}
   \toprule
   \multicolumn{5}{l}{\textbf{Hexagonal close-packed (hcp)}}                                               \\
   \midrule
   Number of possible motifs ($n_\t{c}^{13}$)                            & $8,192$ & $1,594,323$ & $67,108,864$ & $1,220,703,125$ \\
   Number of unique compositions $\l[\frac{n_\t{c}(n_\t{c}+11)!}{12!(n_\t{c}-1)!}\r]$ & $26$    & $273$      & $1,820$        & $9,100$       \\
   Number of distinct motifs $(N_\t{dm})$  & $872$   & $140,022$  & $5,700,480$  & $102,656,875$  \\
   \bottomrule
 \end{tabularx}
  \caption{\label{tbl:Polya}\textbf{Counting of distinct chemical motifs.} Three different crystal lattices with $n_\t{c} = 2, 3, 4,$ or 5 chemical elements were considered. Notice the increasing gap between WC-like representations and the complete chemical motif representation. For a five-element hcp system the motif representation results in an increase of four orders of magnitude in the complexity of the chemical representation.}
\end{table*}

Consider a system with $n_\t{c}$ chemical elements in a crystal structure in which atoms have $N_\t{a}$ \gls{1NN}. Out of the $n_\t{c}^{N_\t{a}+1}$ possible chemical motifs that can be constructed, only a select few are distinct. To analytically count the number of distinct chemical motifs we apply the Polya's \textit{pattern inventory} formula (\cref{eq:pattern_inventory}), which stems from Polya's enumeration theory (see the Methods section ``\hyperref[sec:Polya]{Polya’s enumeration theory}''). This approach is a general mathematical formalism, based on group theory, for counting the number of distinct colorings of objects under the action of a permutation group. When adapting Polya's theory to the counting of distinct chemical motifs, the objects being counted are only the 1CPs because the role of the central atom (fig.~\rref{fig:fig_framework}{a}) is trivial to account for as a simple multiplicative factor given by the number of atom types $n_\t{c}$ (represented here by the number of different colors). The symmetry of the 1CP is defined by the crystal lattice, which also defines the permutation group under consideration.

We illustrate this approach by applying it to \gls{fcc}, \gls{bcc}, and \gls{hcp} three-element ($n_\t{c}=3$) systems. The outcome of Polya's pattern inventory formula (\cref{eq:pattern_inventory}) is a polynomial in which the coefficients indicate the chemical composition of the \gls{1CP}, while the prefactors are the number of distinct \gls{1CP}s with that same composition. While these polynomials are given in Supplementary Section 2, the information contained in them is better represented in a ternary diagram (\cref{fig:fig_pattern_inventory}) that indicates the chemical composition of the \gls{1CP} along with the corresponding number of distinct 1CPs with that composition. The number of distinct 1CPs for a fixed chemical composition quantifies the degeneracy of the WC representation for that 1CP. For example, for the three-element fcc system (fig.~\rref{fig:fig_pattern_inventory}{a}) there are 768 distinct equiatomic 1CPs, which all have the same WC parameters given a specific central atom type. Summing the number of distinct motifs across this compositional diagram (i.e., all of the 91 possible 1CP compositions) results in a total of only $3\times 12,111 = 36,333$ distinct chemical motifs, out of the $3^{13} = 1,594,323$ possible ones.

The enumeration of distinct motifs reveals the incomplete description of \gls{WC}-like parameters, which may result in the misleading characterization of the diversity of chemical motifs in physical systems. This problem becomes particularly alarming for chemically complex materials, such as high-entropy alloys and ceramics. In table~\ref{tbl:Polya} we have applied our approach to up to $n_\t{c} = 5$ chemical elements. Consider, for example, that for a five-element \gls{hcp} system there are only $9,100$ distinct chemical motif compositions, but more than 100 million distinct chemical motifs: a difference of \textit{four orders of magnitude} in the complexity of the chemical representation.

\subsection{Classifying chemical motifs with machine learning}\label{sec:micro_id}

\begin{figure}
  \centering
  \includegraphics[width=\columnwidth]{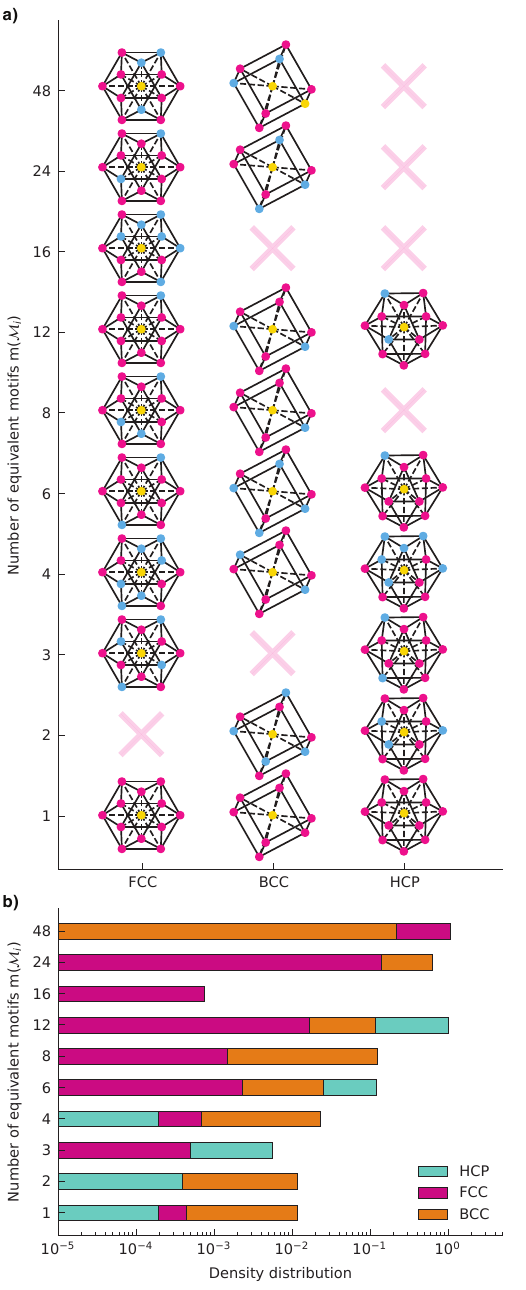}
  \caption{\label{fig:fig_synthetic_multiplicities}
  \textbf{Degeneracy of chemical motifs.} The total number of motifs equivalent to $\mathcal{M}_i$ is given by $m(\mathcal{M}_i)$. \textbf{a)} Illustration of one motif for each of the possible values of $m(\mathcal{M}_i)$ observed in the three-element systems of fig.~\ref{fig:fig_pattern_inventory}. \textbf{b)} Fraction of distinct motifs with a given value of $m(\mathcal{M}_i)$.}
\end{figure}

While Polya's theory enables the counting of distinct chemical motifs, it is not capable of classifying an unidentified motif, which is a fundamental component of the framework illustrated in fig.~\rref{fig:fig_framework}{c}. Classification of an unknown chemical motif requires finding which distinct motif is equivalent to the unknown motif. However, rigorously establishing this equivalency requires the determination of graph isomorphisms, which is a computational task not typically solvable within polynomial time. Here we circumvent this computational limitation by employing a randomly-initialized E(3)-equivariant graph neural network\autocite{rackers_recipe_2023} (E(3)-GNN, see the Methods section ``\hyperref[sec:Network]{\texorpdfstring{E(3)\t{-}}-equivariant graph neural network}''). GNNs are often capable of creating representations that capture intra-graph relationships and topology, effectively distinguishing between graph structures\autocite{xu2019powerful}. This capability can also be understood by the similarity between GNNs' message-passing algorithm to the Weisfeiler-Lehman test for graph isomorphism \autocite{Leman2018THERO, morris_weisfeiler_2019}.

Our E(3)-GNN employs a graph convolutional neural network on the 1CP, where nodes represent atoms and edges represent the connections between them. The network processes this data through hidden layers where graph features undergo transformations adhering to the principles of irreducible representations of 3D rotations and spatial inversion; in the end, a fingerprint $\v{z}_i$ that encodes the chemical motif $\mathcal{M}_i$ is generated. With this neural network architecture any two \textit{equivalent} chemical motifs (i.e., motifs that can be related to each other by a E(3) symmetry operation) are guaranteed to have the same fingerprint; or, if $\mathcal{M}_i \sim \mathcal{M}_j$ then $\v{z}_i = \v{z}_j$.

\begin{figure*}[!bt]
  \centering
 \includegraphics[width=\textwidth]{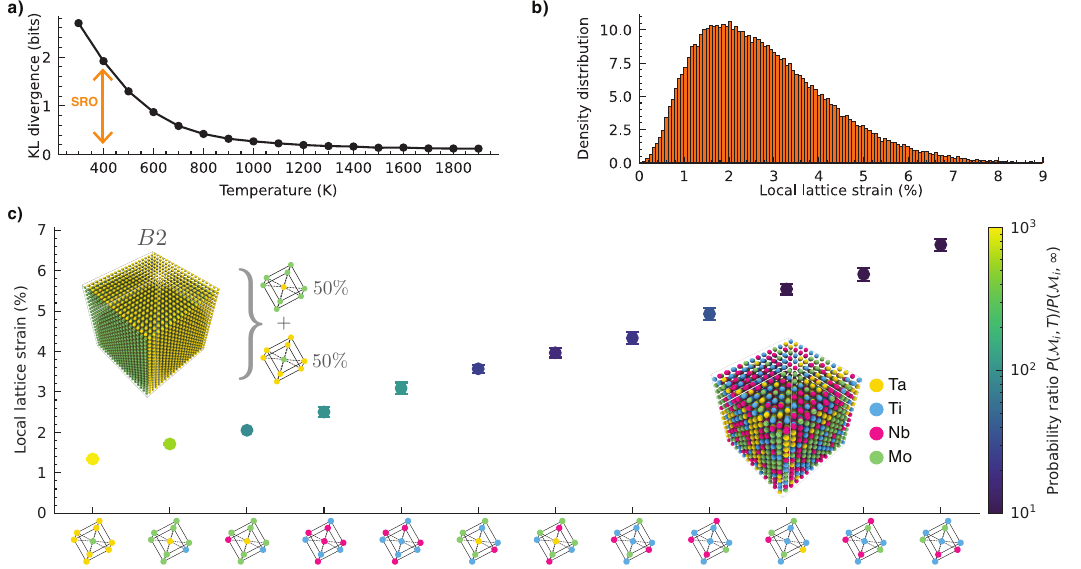}
  \caption{\label{fig:fig_local_lattice_displacements} \textbf{Quantification of chemical short-range order in bcc MoTaNbTi.} \textbf{a)} The Kullback-Leibler divergence ($D_{\t{KL}}$) is a proper information-theoretic quantification of chemical short-range order, i.e., the difference between the probability of observing a local chemical motif in thermal equilibrium $P(\mathcal{M}_i, T)$ versus in a random solid solution $P(\mathcal{M}, \infty)$. \textbf{b)} Without the local chemical motif representation there is no atomic-level granularity in understanding the distribution of local lattice strain across the system. \textbf{c)} Association of representative chemical motifs with their corresponding local lattice strain and probability $P(\mathcal{M}_i, T)$ at $T = 300\,\t{K}$. The inset shows that the three motifs with lowest local strain are variations of motifs observed in a B2-ordered alloy.}
\end{figure*}

One nuanced point of this approach is that it does not guarantee that any two \textit{distinct} chemical motifs will have different fingerprints, i.e., if $\mathcal{M}_i \nsim \mathcal{M}_j$ then it is not guaranteed that $\v{z}_i \neq \v{z}_j$. Thus, validation is required. Here we accomplish this by evaluating the pattern inventories in table \ref{tbl:Polya} with E(3)-GNN as follows. First we create a data set with all possible 1CPs for a given crystal lattice and number of chemical elements, then we compute the fingerprint of each 1CP with E(3)-GNN. A grouping algorithm is employed to cluster all equivalent 1CPs, from which the number of distinct 1CPs with the same composition can be counted. This \gls{ML}-obtained pattern inventory matches exactly each of the analytically-obtained pattern inventories in table \ref{tbl:Polya}, confirming that our framework is able to differentiate between any possible chemical motifs in these systems. Notice how this approach can quickly become computationally intractable: a five-element hcp system requires the creation of 1.2 billion graphs. This is addressed in Supplementary Section 3, where we discuss how to significantly reduce the computational cost by validating the network's expressivity solely on a symmetrically-complete subset of the data set, resulting in orders of magnitude savings.

\subsection{Information-theoretic quantification of short-range order}\label{sec:pop_density}

The probability of observing a distinct motif $\mathcal{M}_i$ at temperature $T$ is
\begin{equation*}
  \label{eq:p_tess}
  P(\mathcal{M}_i, T) = \frac{N(\mathcal{M}_i,T)}{\sum\limits_{j=1}^{N_\t{dm}} N(\mathcal{M}_j,T)},
\end{equation*}
where $N(\mathcal{M}_i,T)$ is the number of times that $\mathcal{M}_i$ is observed in the thermally equilibrated system at temperature $T$, and the sum in the denominator is over all $N_\t{dm}$ distinct motifs $\mathcal{M}_j$. A random SS (indicated here by $T = \infty$) is such that $N(\mathcal{M}_i,\infty)$ is proportional to the total number of motifs equivalent to $\mathcal{M}_i$, given by $m(\mathcal{M}_i)$. For example, fig.~\rref{fig:fig_synthetic_multiplicities}{a} illustrates one motif for each of the possible values of $m(\mathcal{M}_i)$ observed in the three-element systems of fig.~\ref{fig:fig_pattern_inventory}. Meanwhile fig.~\rref{fig:fig_synthetic_multiplicities}{b} shows the fraction of distinct motifs with a given value of $m(\mathcal{M}_i)$.

Thermal effects in a real SS induce a trade-off between enthalpy and entropy that favors low-energy chemical motifs\autocite{yang_coordination_2008, polak_local_2003, clapp_atomic_1971,sheriff2023quantifying}. This deviation from randomness --- namely, chemical SRO --- can be quantified with a proper information-theoretic measurement of the difference between $P(\mathcal{M},T)$ and $P(\mathcal{M},\infty)$, known as the \gls{KL} divergence\autocite{joyce_kullback-leibler_2011}:
\begin{flalign}
  \label{eq:kl}
  D_{\t{KL}} \Big[ P(\mathcal{M},T) \; || \; P(\mathcal{M},\infty) \Big] = && \nonumber\\ 
  \sum_{i=1}^{N_\t{dm}} P(\mathcal{M}_i, &T) \; \log_2 \bigg[\f{P(\mathcal{M}_i, T)}{P(\mathcal{M}_i, \infty)} \bigg].
\end{flalign}

To illustrate the capabilities of this approach we have evaluated $P(\mathcal{M}_i,T)$ through \gls{MC} simulations for the \gls{bcc} high-entropy alloy MoTaNbTi across a wide range of temperatures (see the Methods section ``\hyperref[sec:mc]{Monte Carlo simulations}''). The KL divergence (fig.~\rref{fig:fig_local_lattice_displacements}{a}) is seen to approach zero at high temperatures, indicating that the probability of observing motifs is converging towards a random SS (i.e., entropy dominated). Conversely, at low temperatures, the \gls{KL} divergence is significantly different from zero, indicating deviations from a random SS due to SRO (i.e., an entropy--enthalpy trade-off). Figure \rref{fig:fig_local_lattice_displacements}{a} shows that eq.~\ref{eq:kl} is a convenient form of summarizing the complex information about all motifs --- provided by $P(\mathcal{M}_i,T)$ --- into a single quantity.

This approach also enables the association of any per-atom property with their corresponding motif and $P(\mathcal{M}_i,T)$. For example, fig.~\rref{fig:fig_local_lattice_displacements}{b} shows the distribution local lattice strains\autocite{owen_lattice_2018, owen_assessment_2017} for the entire system (see the Methods section ``\hyperref[sec:local_lattice_distortion]{Local lattice strains}''), with no atomic-scale granularity in understanding. Meanwhile, using our approach we obtain fig.~\rref{fig:fig_local_lattice_displacements}{c}, where representative chemical motifs are associated with their corresponding local lattice strain. In this figure it can be seen that lower strains are associated with motifs that are observed much more often in thermal equilibrium when compared to a random SS, as measured by the probability ratio $P(\mathcal{M}_i, T) / P(\mathcal{M}_i, \infty)$ that is part of eq.~\ref{eq:kl}.

The inset in fig.~\rref{fig:fig_local_lattice_displacements}{c} further illustrates the nuanced characterization of SRO provided by our approach. It is commonly accepted that chemical SRO is the precursor of ordered structures (or precipitates), such as the B2 structure shown in fig.~\rref{fig:fig_local_lattice_displacements}{c} (inset on the left). In this figure we quantify this concept by showing that the two motifs associated with the TaMo B2 exhibit lower than average local lattice strain, while being significantly more frequent in the thermally equilibrated system than in a random SS. Notice that because the system is still in the SS phase (fig.~\rref{fig:fig_local_lattice_displacements}{c}, inset on the right), the B2 motif with Ta at the center is more often observed in ``defected'' states (i.e., with one Ti or Nb atom substituting the site of a Mo) than in their ideal configuration expected from the B2 ordered structure. Supplementary Section 5 shows the decomposition of other ordered crystal structures into local chemical motifs.

\subsection{Motifs dissimilarity}\label{sec:dissim}

The relative probability of motifs (eq.~\ref{eq:kl}) is not a complete characterization of SRO because it does not contain any information about the spatial distribution of these motifs. This missing information can be provided by rigorously-defined correlation functions\autocite{sheriff2023quantifying} between a motif $\mathcal{M}_i$ and other motifs at a distance $r$ from $\mathcal{M}_i$:
\begin{equation}
  \label{eq:phi_i}
  \phi_i(r,T) = 1 - 2 \, \big< d_{ij} \big>_{|\v{r}_i-\v{r}_j| = r},
\end{equation}
where $d_{ij}$ is a dissimilarity measure between motifs $\mathcal{M}_i$ and $\mathcal{M}_j$ (i.e., it quantifies how different these two motifs are), with $\l< \ldots \r>_{|\v{r}_i-\v{r}_j| = r}$ being an average that includes only motifs $\mathcal{M}_j$ (located at $\v{r}_j$) at a distance $r$ from $\mathcal{M}_i$ (located at $\v{r}_i$). While the framework developed in ref.~\cite{sheriff2023quantifying} is still applicable, the dissimilarity measure $d_{ij}$ needs to be extended to account for arbitrary lattice geometries and number of elements.

\begin{figure*}[!bt]
  \centering
 \includegraphics[width=\textwidth]{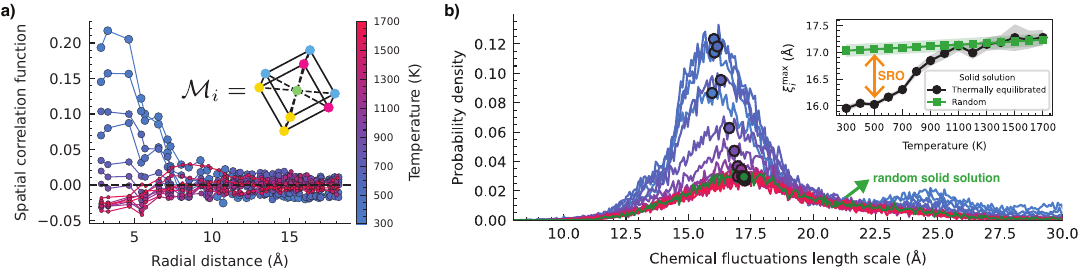}
  \caption{\label{fig:fig_characterization}\textbf{Length scale of chemical fluctuations in bcc MoTaNbTi.} \textbf{a)} Spatial correlation function for a representative chemical motif. Higher temperatures reduce the amount of spatial correlation between motifs. \textbf{b)} Probability distribution of the length scale of chemical fluctuations ($\xi_i$). Inset shows the temperature dependence of the maximum of the probability distribution, where it can be seen that the distribution of chemical fluctuations converge towards the distribution of a random SS (i.e., entropy dominated) at high temperatures\rev{.}}
\end{figure*}

Here we generalize the definition of $d_{ij}$ (eq. 4 in ref.~\cite{sheriff2023quantifying}) by rewriting it as the sum of three separate terms:
\begin{equation}\label{eq:diss}
  d_{ij}  = w_1 \cdot d_{ij}^\t{cat} + w_2 \cdot d_{ij}^\t{com} + w_3 \cdot d_{ij}^\t{dmo},
\end{equation}
where the weights $w_1$, $w_2$, and $w_3$ govern the importance of each term, which are all normalized to fall within the closed interval $[0,1]$. The definition of each dissimilarity component in eq.~\ref{eq:diss} is described next, with the support of Supplementary Figure 3 as a visual guide to the calculation of each component.

The first term ($d_{ij}^\t{cat}$) captures the dissimilarity between the central atoms
\begin{equation}\label{eq:dcat}
  d_{ij}^\t{cat}=  1 - \delta_{\tau_i\tau_j},
\end{equation}
where $\tau_i $ is the central atom type of $\mathcal{M}_i$, and $\delta$ is the Kronecker delta.

The second term ($d_{ij}^\t{com}$) represents the dissimilarity between the chemical compositions of the \gls{1CP}s:
\begin{equation*}
  d_{ij}^\t{com} = \l \lVert  \v{k}_i - \v{k}_j \r \rVert_2,
\end{equation*}
where $\v{k}_i$ are the Cartesian coordinates obtained from the barycentric coordinates of a $(n_\t{c}-1)$-simplex (see the Methods section ``\hyperref[sec:simplex]{Simplex and barycentric coordinates}''). For example, for a ternary alloy the motifs at different vertices of the composition triangle are separated by a dissimilarity distance of one (see ``Chemical composition space" in Supplementary Figure 3).

Finally, $d_{ij}^\t{dmo}$ represents the dissimilarity between distinct motifs with the same 1CP chemical composition:
\begin{equation}\label{eq:dijmo}
  d_{ij}^\t{dmo} = \l \lVert  \v{z}_i  -  \v{z}_j \r \Vert_2 \times \frac{1}{M},
\end{equation}
where $\v{z}_i$ is the E(3)-GNN embedding of the $\mathcal{M}_i$ graph, and $M$ is a normalization factor set to the maximum $L_2$ distance among all possible motifs. 

The weights ($w_1$, $w_2$, and $w_3$) for each dissimilarity component are chosen to be proportional to the number of chemical bonds associated with their corresponding structures in the motif:
\begin{equation*}
  \v{w} =
  \begin{bmatrix}
    w_1 \\ w_2 \\ w_3
  \end{bmatrix}
  = 
  \frac{1}{N_\t{a} +2N_\t{b}} \cdot
   \begin{bmatrix} 
    N_\t{a} \\ N_\t{b} \\  N_\t{b}
  \end{bmatrix} ,
\end{equation*}
where $N_\t{a}$ is the number of atoms in the \gls{1CP}, and $N_\t{b}$ is the number of bonds within the \gls{1CP}. For example, the \gls{fcc} crystal structure has $N_\t{a} = 12$ and $N_\t{b} = 24$, while in bcc $N_\t{a} = 8$ and $N_\t{b} = 12$.

Using this generalized approach we have completed the characterization of SRO in MoTaNbTi, which was initiated in fig.~\ref{fig:fig_local_lattice_displacements}. Figure \rref{fig:fig_characterization}{a} shows the spatial correlation function for a representative motif in this system. With correlation functions such as this one it is possible to evaluate the length scale\autocite{sheriff2023quantifying} ($\xi_i$) of chemical fluctuations for each motif $\mathcal{M}_i$, which is an important materials parameter in the understanding of various chemistry--microstructure relationships\autocite{xie_percolation_2021,varvenne_theory_2016}. Figure \rref{fig:fig_characterization}{b} shows the probability distribution of the length scale of chemical fluctuations, where it can be seen that the effect of SRO is to decrease the average length scale. Figure \ref{fig:fig_characterization} also shows that the distribution of chemical fluctuations converge towards the distribution of a random SS (i.e., entropy dominated) at high temperatures, similarly to what was observed in fig.~\rref{fig:fig_local_lattice_displacements}{a} for the probability distribution of chemical motifs.

\rev{Notice that $d_{ij}^\t{dmo}$ (eq.~\ref{eq:dijmo}) directly depends on the E(3)-GNN embedding, which makes the correlation functions implicitly depend on the model parameters. In Supplementary Section 7 we demonstrate that the chemical fluctuation length scale is robust against different random initializations of the E(3)-GNN. The authors believe that this robustness originates from the correlation function $C_i(r,T) = \phi_i(r,T)  - \phi^0_i(r,T)$ definition\autocite{sheriff2023quantifying}, which is relative to a ``baseline'' $\phi^0_i(r,T)$ that is also influenced by the random initialization:
\[
\phi^0_i(r,T) = 1 - 2 \, \big< d_{ij} \big>_{P(\mathcal{M},T)},
\]
where $\l< \ldots \r>_{P(\mathcal{M},T)}$ indicates an average evaluated with motifs $\mathcal{M}_j$ randomly sampled from the thermally equilibrated distribution $P(\mathcal{M},T)$.}

\section{Discussion}

The generalized framework presented here characterizes chemical fluctuations in arbitrary crystal lattices with any number of chemical elements, with a rigorous mathematical foundation for the generalization being provided. The framework culminates in a reduced representation of the chemical space (Supplementary Figure 3), and an information-theoretic quantification of SRO. Analytical results using group theory demonstrate that this approach eliminates degeneracies present in other representations of SRO (fig.~\rref{fig:fig_framework}{b}). \rev{Our framework can identify motifs in atomistic data with a computational speed of $1.2\times 10^6$ atoms per hour in a single CPU core with a Apple Silicon M1 processor, or $63 \times 10^6$ atoms per hour on a NVIDIA V100s GPU. While this is indeed more computationally expensive than computing first nearest-neighbors Warren-Cowley parameters, the computational cost is still negligible for typical large-scale atomistic simulations with tens of millions of atoms.} The application of this framework to arbitrary crystal structures can be automated by employing symmetry finder algorithms\autocite{csardi_igraph_2023} to determine the symmetry group of chemical motifs. It is important to consider the neural-network expressivity and memory constraints when expanding the scope of this method, especially when working with high-entropy materials. In Supplementary Section 3 we provide a series of strategies for validating the expressivity on a symmetrically-complete subset of the data set, which enables the application of this approach to systems with at least 1.2 billion motifs. 

\rev{The approach introduced here is complete for the first-coordination environment of an atom, while the WC parameters are not (fig.~\rref{fig:fig_framework}{b}). Yet, our approach is \textit{not complete} beyond the first-coordination environment, and this subtle point warrants some discussion as it remains a long-standing challenge in the field. The incompleteness of our approach beyond first-neighbors is due to the fact that the full set of correlation functions (eq.~\ref{eq:phi_i}) is still not enough to completely reconstruct the chemical state of the system. Fundamentally, this originates from the aggregation of the contribution of all motifs $\mathcal{M}_j$ to the correlation function of motif $\mathcal{M}_i$ in eq.~\ref{eq:phi_i}. A complete description would require pairwise correlation functions (i.e., individual correlations between all motif pairs), and possibly higher-order correlation functions (e.g., between motif triplets), which might be computationally impractical to evaluate. Perhaps more importantly than computational feasibility: the authors are not aware of experimental measurements on this complete set of correlation functions, or any suggestion in the literature of how they affect materials properties. Meanwhile, the set of correlation functions evaluated here (eq.~\ref{eq:phi_i}) is in principle physically equivalent to approaches currently in use to evaluate SRO length scale with electron microscopy\autocite{xu_correlating_2023,xu_determination_2023}, and have been historically employed to evaluate experimental observables such as scattering cross sections and susceptibilities in statistical physics\autocite{xu_correlating_2023,xu_determination_2023,correlated_disorder,KL_correlation}.

A discussion is also warranted regarding the interpretability of our approach beyond the first-coordination environment when compared to WC parameters. By aggregating (eq.~\ref{eq:diss}) the contribution of the central atom chemical species ($d_{ij}^\t{cat}$, eq.~\ref{eq:dcat}) with the other contributions (namely, 1CP chemical composition $d_{ij}^\t{com}$ and motif structure $d_{ij}^\t{dmo}$) in the calculation of the correlation function (eq.~\ref{eq:phi_i}) we end up with a description beyond the first-coordination environment that is less interpretable than WC parameters, which describe the variation in chemical composition with distance in a physically intuitive manner. This reduced interpretability is traded by an increase in the physical fidelity of the evaluation of SRO length scales by the inclusion of effects beyond chemical composition (i.e., $d_{ij}^\t{com}$ and $d_{ij}^\t{dmo}$).

Extending our current framework to account for long-range order would be a natural continuation of the work presented here, because this would allow the investigation of phenomena where ordered compounds and solid solutions are both present (e.g., precipitation hardening). This could be accomplished, for example, with motif-node based graphs\autocite{banjade2021structure,wang2023motif}, where each node in the graph describes an entire chemical motif, or with variational autoencoder based order parameters\autocite{yin_neural_2021}. In Supplementary Section 8 we show that the chemical fluctuation length scale (fig.~5b) shows variations that are compatible with peaks in the specific heat of CrCoNi that have been attributed to transitions to long-range order (i.e., a order-disorder transition).
}

The capabilities developed here facilitate the evaluation of chemistry--microstructure relationships that will be valuable for materials theory and experiments alike. For example, this approach is useful for augmenting the visualization of large-scale atomistic simulations\autocite{cao_maximum_2022, yin_atomistic_2021, li_strengthening_2019,ghosh_short-range_2022,kostiuchenko_impact_2019, du_chemical_2022} or experimental imaging at the atomic scale\autocite{zhou_atomic-scale_2022, zhang_short-range_2020, chen_direct_2021, xu_determination_2023, hsiao_data-driven_2022, wang_chemical_2022,xu_correlating_2023,hu_application_2022, moniri_three-dimensional_2023, li_machine_2023, li_convolutional_2021}, leading to better characterization of chemical SRO and its connection with physical properties (e.g., figs.~\rref{fig:fig_characterization}{b} and ~\rref{fig:fig_local_lattice_displacements}{c}). Our approach could also better inform the chemistry during the development of machine learning interatomic potentials for chemically-complex systems\autocite{cao2024capturing, freitas_machine-learning_2022, lopanitsyna_modeling_2023, willatt_feature_2018, darby_compressing_2022}, which is currently a challenge for the state-of-the-art in the field. The results presented here demonstrate how data science and machine learning can be employed to uncover chemical complexity in large atomic-scale data sets, and transform these findings into quantities of relevance for the physical modeling of these materials.

\clearpage
\twocolumn[
  \begin{center}
    \Large
    \textbf{Methods}
  \end{center}
]

\subsection{Polya's enumeration theory}\label{sec:Polya}

Consider a system with $n_\t{c}$ chemical elements in a specified crystal structure, which defines the 1CP. We define $S$ to be the set of vertices of a 1CP, $C$ to be the set of possible atom types (i.e., chemical elements), and $G$ to be a group isomorphic to the \gls{1CP} symmetry group.
The \gls{1CP} pattern inventory is given by Polya's enumeration theory\autocite{polya_kombinatorische_1937}:
\begin{equation}\label{eq:pattern_inventory}
  P_G\l(\sum_{c \in C}w(c),\sum_{c \in C}w(c)^2,...\,,\sum_{c \in C}w(c)^d\r),
\end{equation}
where $P_G$ is the cycle index polynomial of $G$, $w(c)$ is the atomic type label of $c \in C$ (e.g.,``Mo'',``Ta'', ``Nb'', or ``Ti'' for MoTaNbTi), and $d=|S|$ is the cardinality of $S$ (i.e., number of elements in the set). The pattern inventories for \gls{fcc}, \gls{hcp}, and \gls{bcc} lattices with three elements are given in Supplementary Section 2. Pattern inventories for arbitrary lattices and number of elements can be evaluated using our \texttt{Polya} Python package\autocite{polya_github}.
 
Equation \ref{eq:pattern_inventory} contains all of the information required to enumerate (i.e., count) the number of distinct 1CPs for each of its possible chemical compositions. For example, the pattern inventory of a ABC ternary hcp system (Supplementary Section 2) allow us to easily read from the polynomial coefficients that there are 1444 distinct 1CPs with chemical composition A$_5$B$_2$C$_5$. Similarly, the total number of distinct 1CPs ($N_\t{dm}/n_\t{c}$, also known in group theory as the number of orbits) can be obtained by setting $w(c) = 1$ for all $c \in C$ in the polynomial:
\begin{equation*}\label{eq:total_ms}
    \frac{N_\t{dm}}{n_\t{c}} = P_G(|C|, ... \,, |C|).
\end{equation*}

\subsection{E(3)-equivariant graph neural network}\label{sec:Network}

Local chemical motifs were converted to graphs in which nodes represent atoms and have as attribute the atomic type (i.e., chemical element) as a one-hot encoding. Graph edges store the direction (unit) vector $\uv{r}_{ij}$ between atoms $i$ and $j$. \rev{Graph embeddings are made invariant to local lattice distortions by remapping atoms to their ideal positions before constructing the graphs.} Each graph is \rev{then} processed through a randomly initialized E(3)-GNN\autocite{rackers_recipe_2023} (implemented using the \texttt{e3nn} package\autocite{geiger_e3nn_2022}) composed of \rev{$\mathbb{E}(3)-$}equivariant convolutions and gates, generating a fingerprint $\v{z}_i \in \mathbb{R}^{4}$ of the chemical motifs $\mathcal{M}_i$. 

Our E(3)-GNN architecture is composed of three convolutions using spherical harmonics $Y_\ell^m\l(\uv{r}_{ij}\r)$ as edge attributes, with degree as large as $\ell_{max}=2$. A total of 10 cosine radial basis functions are employed, with ranges evenly spread from zero to three distance units (measured as multiples of the nearest-neighbor distance). The final output length is four, i.e., $\v{z}_i \in \mathbb{R}^{4}$. The hidden-layers irreducible representation, in the \texttt{e3nn} notation, were set to $3\times2o+3\times2e+3\times1o+3\times1e+3\times0o+3\times0e$. Equivariant neural networks have been shown to be capable symmetry compilers, enabling the representation of important features of atomic geometry even with random initialization\autocite{smidt_finding_2021}. Our choice here of randomly initializing the network\rev{, and of not training it,} is meant to maximize the influence of each weight. This is also supported by the agreement between the pattern inventories obtained with this computational approach and the analytical results using Polya's enumeration theory.

\subsection{Monte Carlo simulations}\label{sec:mc}

Monte Carlo simulations using the machine learning moment tensor potential\autocite{shapeev_moment_2016} from ref.~\cite{zheng_multi-scale_2023} were employed to sample the thermal-equilibrium chemical configurations for the \gls{bcc} MoTaNbTi high-entropy alloy. Starting configurations were composed of $13 \times 13 \times 13$ chemically-random supercells at the equiatomic composition. The simulations were performed for a total of 30 Monte Carlo steps per atom, with acceptance probabilities based on the Metropolis-Hastings algorithm\autocite{robert_metropolishastings_2004,metropolis_equation_2004}. Periodic boundary conditions were enforced in all dimensions.

Simulations were carried out for temperatures ranging from 300\,K to 1700\,K in 100\,K increments; the lattice parameter at each temperature accounted for thermal expansion. A total of 20 independent simulations were performed for each temperature, and only the final configuration of each simulation was employed in our analyses, which resulted in a data set of 87,880 motifs per temperature. The nearest-neighbor Warren--Cowley parameters of this system (calculated using our Ovito\autocite{stukowski_visualization_2009} \texttt{WarrenCowleyParameters} Python modifier \rev{(\href{https://github.com/killiansheriff/WarrenCowleyParameters}{github.com/killiansheriff/WarrenCowleyParameters})}) is shown in Supplementary Section 4 as a function of temperature.

\subsection{Local lattice strain}\label{sec:local_lattice_distortion}

The final configuration of each Monte Carlo simulation was relaxed with fixed simulation box dimensions, and the local lattice strain of each atom $n$ was evaluated:
\begin{equation*}
\delta_n(T) = \frac{\lVert \mathbf{r}_n^{\text{f}} - \mathbf{r}_n^{\text{i}} \rVert_2}{a_{\text{NN}}(T)},
\end{equation*}
where $\mathbf{r}_n^{\text{f}}$ is the atom's position after relaxation, $\mathbf{r}_n^{\text{i}}$ is the position in before relaxation (i.e., in the ideal bcc structure), $\lVert \ldots \rVert_2$ is the $L^2$ norm, and $a_\text{NN}(T)$ is the nearest-neighbor distance at temperature $T$.

\subsection{Simplex and barycentric coordinates}\label{sec:simplex}

A simplex \autocite{munkres_elements_1993} is a generalization of triangles (2-simplex) and tetrahedra (3-simplex) to higher dimensions. It represents the simplest possible polytope in any given dimension. Mathematically, the standard $n$-simplex is defined as the subset of $\mathbb{R}^{n+1}$ given by
\begin{equation*}
    \resizebox{\hsize}{!}{$\nabla^n = \l\{\v{t} \in \mathbb{R}^{n+1} : \sum_{i=1}^{(n+1)} t_i = 1 \land t_i\geq 0 \t{ for } i = 1,...,(n+1) \r\}$}.
\end{equation*}
A barycentric coordinate system specifies the location of a point with respect to a simplex. In our case, the barycentric coordinates correspond to the chemical composition of a \gls{1CP}. For example, consider a motif $\mathcal{M}_j$ and its associated 1CP. The barycentric coordinates for the composition of this 1CP are
\[
  \v{\lambda}_j =\frac{1}{N_\t{a}}\l(N^{(1)},N^{(2)}, \ldots, N^{(n_\t{c})}\r),
\]
where $N_\t{a}$ is the number of atoms in the 1CP, $N^{(i)}$ is the number of atoms of type ``$i$'', and $n_\t{c}$ is the number of chemical elements in the system. While the barycentric coordinates are a convenient form to express the chemical composition, they can be converted into the more conventional Cartesian coordinates $\v{k}_j \in \mathbb{R}^{(n_\t{c}-1)}$ with the following equation:
\begin{equation*}
    \v{k}_j =  \frac{1}{N_\t{a}} \sum_{i=1}^{n_\t{c}} N^{(i)} \v{v}_i ,
\end{equation*}
where $\v{v}_i$ are vertices of the $(n_\t{c}-1)$-simplex (in Cartesian coordinates) . The numerical implementation of $n$-simplex objects can be found in our \texttt{Simplex} Python package\autocite{simplex_github}.

\subsection{Data availability}
The data that support the findings of this study are available from the corresponding authors upon reasonable request.

\subsection{Code availability}
\rev{
The software for chemical motif identification and SRO quantification can be found in our \texttt{ChemicalMotifIdentifier} Python package\autocite{cmi_github} (\href{https://github.com/killiansheriff/ChemicalMotifIdentifier}{github.com/killiansheriff/ChemicalMotifIdentifier}). Our figure style is implemented in \texttt{LovelyPlots}\autocite{lovelyplots} under the \texttt{paper} style. For convenience, we have compiled the list of all Python packages developed in this work in a GitHub repository list\autocite{repo_list} (\href{https://github.com/stars/killiansheriff/lists/chemical-motif-characterization}{github.com/stars/killiansheriff/lists/chemical-motif-characterization}), which includes our \texttt{ChemicalMotifIdentifier}\autocite{cmi_github}, \texttt{Polya}\autocite{polya_github}, \texttt{Simplex}\autocite{simplex_github}, \texttt{NShellFinder}\autocite{nshell_github} and \texttt{WarrenCowleyParameters}\autocite{wc_github} Python packages.
}

\subsection{Author contributions}

K.S., Y.C., and R.F. conceived the project. K.S. performed all calculations. All authors contributed to the interpretation of the results. K.S. and R.F. prepared the manuscript, which was reviewed and edited by all authors. Project administration, supervision, and funding acquisition was performed by R.F.

\subsection{Acknowledgments}

This work was supported by the MathWorks Ignition Fund, MathWorks Engineering Fellowship Fund, and the Portuguese Foundation for International Cooperation in Science,
Technology and Higher Education in the MIT--Portugal Program. We were also supported by the Research Support Committee Funds from the School of Engineering at the Massachusetts Institute of Technology. This work used the Expanse supercomputer at the San Diego Supercomputer Center through allocation MAT210005 from the Advanced Cyberinfrastructure Coordination Ecosystem: Services \& Support (ACCESS) program, which is supported by National Science Foundation grants \#2138259, \#2138286, \#2138307, \#2137603, and \#2138296, and the Extreme Science and Engineering Discovery Environment (XSEDE), which was supported by National Science Foundation grant number \#1548562.

\subsection{Competing interests}

The authors declare no competing interests.

\clearpage
\printbibliography[heading=bibnumbered,title={References}]

@article{banjade2021structure,
	title        = {Structure motif–centric learning framework for inorganic crystalline systems},
	author       = {Huta R. Banjade  and Sandro Hauri  and Shanshan Zhang  and Francesco Ricci  and Weiyi Gong  and Geoffroy Hautier  and Slobodan Vucetic  and Qimin Yan},
	year         = 2021,
	journal      = {Sci. Adv.},
	volume       = 7,
	doi          = {10.1126/sciadv.abf1754},
	journal_full = {Science Advances}
}

@misc{cao2024capturing,
	title        = {Capturing short-range order in high-entropy alloys with machine learning potentials},
	author       = {Yifan Cao and Killian Sheriff and Rodrigo Freitas},
	year         = 2024,
	archiveprefix = {arXiv},
	eprint       = {2401.06622},
	primaryclass = {cond-mat.mtrl-sci}
}

@article{cao_maximum_2022,
	journal      = {Sci. Adv.},
	author = {Penghui Cao },
    title = {Maximum strength and dislocation patterning in multi–principal element alloys},
    volume = {8},
    number = {45},
    pages = {eabq7433},
    year = {2022},
    doi = {10.1126/sciadv.abq7433},
}

@article{chen_direct_2021,
	title        = {Direct observation of chemical short-range order in a medium-entropy alloy},
	author       = {Chen, Xuefei and Wang, Qi and Cheng, Zhiying and Zhu, Mingliu and Zhou, Hao and Jiang, Ping and Zhou, Lingling and Xue, Qiqi and Yuan, Fuping and Zhu, Jing and Wu, Xiaolei and Ma, En},
	journal      = {Nature},
	volume       = 592,
	pages        = {712--716},
	doi          = {10.1038/s41586-021-03428-z},
	date         = {2021-04},
	journal_full = {Nature}
}

@article{clapp_atomic_1971,
	title        = {Atomic {Configurations} in {Binary} {Alloys}},
	author       = {Clapp, Philip C.},
	journal      = {Phys. Rev. B},
	volume       = 4,
	pages        = {255--270},
	doi          = {10.1103/PhysRevB.4.255},
	date         = {1971-07},
	journal_full = {Physical Review B}
}

@article{correlated_disorder,
	title        = {The crystallography of correlated disorder},
	author       = {Keen, David A. and Goodwin, Andrew L.},
	year         = 2015,
	journal      = {Nature},
	publisher    = {Nature Publishing Group UK London},
	volume       = 521,
	number       = 7552,
	pages        = {303--309},
	doi          = {10.1038/nature14453},
	journal_full = {Nature}
}

@article{coury_origin_2023,
	title        = {On the origin of diffuse intensities in fcc electron diffraction patterns},
	author       = {Coury, Francisco Gil and Miller, Cody and Field, Robert and Kaufman, Michael},
	year         = 2023,
	journal      = {Nature},
	volume       = 622,
	pages        = {742--747},
	doi          = {10.1038/s41586-023-06530-6},
	journal_full = {Nature}
}

@article{cowley_approximate_1950,
	title        = {An {Approximate} {Theory} of {Order} in {Alloys}},
	author       = {Cowley, J. M.},
	journal      = {Phys. Rev.},
	volume       = 77,
	pages        = {669--675},
	doi          = {10.1103/PhysRev.77.669},
	date         = {1950-03},
	journal_full = {Physical Review}
}

@misc{csardi_igraph_2023,
	title        = {igraph},
	author       = {Csárdi, Gábor and Nepusz, Tamás and Horvát, Szabolcs and Traag, Vincent and Zanini, Fabio and Noom, Daniel},
	publisher    = {Zenodo},
	doi          = {10.5281/zenodo.8143064},
	date         = {2023-07}
}

@article{darby_compressing_2022,
	title        = {Compressing local atomic neighbourhood descriptors},
	author       = {Darby, James P. and Kermode, James R. and Csányi, Gábor},
	year         = 2022,
	journal      = {npj Comput Mater},
	volume       = 8,
	doi          = {10.1038/s41524-022-00847-y},
	journal_full = {npj Computational Materials}
}

@article{ding_tunable_2018,
	title        = {Tunable stacking fault energies by tailoring local chemical order in {CrCoNi} medium-entropy alloys},
	author       = {Ding, Jun and Yu, Qin and Asta, Mark and Ritchie, Robert O.},
	journal      = {Proc. Natl. Acad. Sci. U.S.A.},
	volume       = 115,
	pages        = {8919--8924},
	doi          = {10.1073/pnas.1808660115},
	date         = 2018,
	journal_full = {Proceedings of the National Academy of Sciences of the United States of America}
}

@article{du_chemical_2022,
	title        = {Chemical domain structure and its formation kinetics in {CrCoNi} medium-entropy alloy},
	author       = {Du, Jun-Ping and Yu, Peijun and Shinzato, Shuhei and Meng, Fan-Shun and Sato, Yuji and Li, Yangen and Fan, Yiwen and Ogata, Shigenobu},
	journal      = {Acta Materialia},
	volume       = 240,
	pages        = 118314,
	doi          = {10.1016/j.actamat.2022.118314},
	date         = {2022-11},
	journal_full = {Acta Materialia}
}

@article{feng_effects_2017,
	title        = {Effects of {Short}-{Range} {Order} on the {Magnetic} and {Mechanical} {Properties} of {FeCoNi}({AlSi})x {High} {Entropy} {Alloys}},
	author       = {Feng, Wenqiang and Qi, Yang and Wang, Shaoqing},
	journal      = {Metals},
	volume       = 7,
	pages        = 482,
	doi          = {10.3390/met7110482},
	date         = {2017-11},
	journal_full = {Metals}
}

@article{freitas_machine-learning_2022,
	title        = {Machine-learning potentials for crystal defects},
	author       = {Freitas, Rodrigo and Cao, Yifan},
	year         = 2022,
	journal      = {MRS Communications},
	volume       = 12,
	pages        = {510--520},
	doi          = {10.1557/s43579-022-00221-5},
	journal_full = {MRS Communications}
}

@article{gault_atom_2021,
	title        = {Atom probe tomography},
	author       = {Gault, Baptiste and Chiaramonti, Ann and Cojocaru-Mirédin, Oana and Stender, Patrick and Dubosq, Renelle and Freysoldt, Christoph and Makineni, Surendra Kumar and Li, Tong and Moody, Michael and Cairney, Julie M.},
	year         = 2021,
	journal      = {Nat Rev Methods Primers},
	volume       = 1,
	doi          = {10.1038/s43586-021-00047-w},
	journal_full = {Nature Reviews Methods Primers}
}

@article{gault_spatial_2010,
	title        = {Spatial {Resolution} in {Atom} {Probe} {Tomography}},
	author       = {Gault, Baptiste and Moody, Michael P. and Geuser, Frederic De and Fontaine, Alex La and Stephenson, Leigh T. and Haley, Daniel and Ringer, Simon P.},
	year         = 2010,
	journal      = {Microsc Microanal},
	volume       = 16,
	pages        = {99--110},
	doi          = {10.1017/S1431927609991267},
	journal_full = {Microscopy and Microanalysis}
}

@misc{geiger_e3nn_2022,
      title={e3nn: Euclidean Neural Networks}, 
      author={Mario Geiger and Tess Smidt},
      year={2022},
      eprint={2207.09453},
      archivePrefix={arXiv},
      primaryClass={cs.LG},
      url={https://arxiv.org/abs/2207.09453}, 
}

@article{george_high-entropy_2019,
	title        = {High-entropy alloys},
	author       = {George, Easo P. and Raabe, Dierk and Ritchie, Robert O.},
	journal      = {Nat Rev Mater},
	volume       = 4,
	pages        = {515--534},
	doi          = {10.1038/s41578-019-0121-4},
	date         = {2019-08},
	journal_full = {Nature Reviews Materials}
}

@article{ghosh_short-range_2022,
	title        = {Short-range order and phase stability of {CrCoNi} explored with machine learning potentials},
	author       = {Ghosh, Sheuly and Sotskov, Vadim and Shapeev, Alexander V. and Neugebauer, Jörg and Körmann, Fritz},
	journal      = {Phys. Rev. Materials},
	volume       = 6,
	doi          = {10.1103/PhysRevMaterials.6.113804},
	date         = {2022-11},
	journal_full = {Physical Review Materials}
}

@article{hsiao_data-driven_2022,
	title        = {Data-driven electron-diffraction approach reveals local short-range ordering in {CrCoNi} with ordering effects},
	author       = {Hsiao, Haw-Wen and Feng, Rui and Ni, Haoyang and An, Ke and Poplawsky, Jonathan D. and Liaw, Peter K. and Zuo, Jian-Min},
	journal      = {Nat Commun},
	volume       = 13,
	doi          = {10.1038/s41467-022-34335-0},
	date         = {2022-11},
	journal_full = {Nature Communications}
}

@article{hu_application_2022,
	title        = {Application of atom probe tomography in understanding high entropy alloys: {3D} local chemical compositions in atomic scale analysis},
	shorttitle   = {Application of atom probe tomography in understanding high entropy alloys},
	author       = {Hu, Rong and Jin, Shenbao and Sha, Gang},
	journal      = {Progress in Materials Science},
	series       = {A {Festschrift} in {Honor} of {Brian} {Cantor}},
	volume       = 123,
	pages        = 100854,
	doi          = {10.1016/j.pmatsci.2021.100854},
	date         = {2022-01},
	journal_full = {Progress in Materials Science}
}

@article{joress_why_2023,
	title        = {Why is {EXAFS} for complex concentrated alloys so hard? {Challenges} and opportunities for measuring ordering with {X}-ray absorption spectroscopy},
	shorttitle   = {Why is {EXAFS} for complex concentrated alloys so hard?},
	author       = {Joress, Howie and Ravel, Bruce and Anber, Elaf and Hollenbach, Jonathan and Sur, Debashish and Hattrick-Simpers, Jason and Taheri, Mitra L. and DeCost, Brian},
	year         = 2023,
	journal      = {Matter},
	volume       = 6,
	pages        = {3763--3781},
	doi          = {10.1016/j.matt.2023.09.010},
	journal_full = {Matter}
}

@incollection{joyce_kullback-leibler_2011,
	title        = {Kullback-{Leibler} {Divergence}},
	author       = {Joyce, James M.},
	booktitle    = {International {Encyclopedia} of {Statistical} {Science}},
	location     = {Berlin, Heidelberg},
	publisher    = {Springer},
	doi          = {10.1007/978-3-642-04898-2_327},
	isbn         = {978-3-642-04898-2},
	date         = 2011,
	editor       = {Lovric, Miodrag}
}

@article{KL_correlation,
	title        = {Network information and connected correlations},
	author       = {Schneidman, Elad and Still, Susanne and Berry, Michael J. and Bialek, William},
	year         = 2003,
	journal      = {Phys. Rev. Lett.},
	publisher    = {APS},
	volume       = 91,
	number       = 23,
	pages        = 238701,
	doi          = {10.1103/PhysRevLett.91.238701},
	journal_full = {Physical Review Letters}
}

@article{kostiuchenko_impact_2019,
	title        = {Impact of lattice relaxations on phase transitions in a high-entropy alloy studied by machine-learning potentials},
	author       = {Kostiuchenko, Tatiana and Körmann, Fritz and Neugebauer, Jörg and Shapeev, Alexander},
	journal      = {npj Comput Mater},
	volume       = 5,
	doi          = {10.1038/s41524-019-0195-y},
	date         = {2019-05},
	journal_full = {npj Computational Materials}
}

@article{Leman2018THERO,
	title        = {{A Reduction of a Graph to a Canonical Form and an Algebra Arising During This Reduction}},
	author       = {Weisfeiler, Boris and Lehman, A. A.},
	year         = 1968,
	journal      = {Nauchno-Technicheskaya Informatsia},
	url          = {https://www.iti.zcu.cz/wl2018/pdf/wl_paper_translation.pdf}
}

@article{li_convolutional_2021,
	title        = {Convolutional neural network-assisted recognition of nanoscale {L12} ordered structures in face-centred cubic alloys},
	author       = {Li, Yue and Zhou, Xuyang and Colnaghi, Timoteo and Wei, Ye and Marek, Andreas and Li, Hongxiang and Bauer, Stefan and Rampp, Markus and Stephenson, Leigh T.},
	journal      = {npj Comput Mater},
	volume       = 7,
	doi          = {10.1038/s41524-020-00472-7},
	date         = {2021-01},
	journal_full = {npj Computational Materials}
}

@misc{li_machine_2023,
      title={Machine learning-enabled tomographic imaging of chemical short-range atomic ordering}, 
      author={Yue Li and Timoteo Colnaghi and Yilun Gong and Huaide Zhang and Yuan Yu and Ye Wei and Bin Gan and Min Song and Andreas Marek and Markus Rampp and Siyuan Zhang and Zongrui Pei and Matthias Wuttig and Jörg Neugebauer and Zhangwei Wang and Baptiste Gault},
      year={2023},
      eprint={2303.13433},
      archivePrefix={arXiv},
      primaryClass={cond-mat.mtrl-sci},
      url={https://arxiv.org/abs/2303.13433}, 
}

@article{li_strengthening_2019,
	title        = {Strengthening in multi-principal element alloys with local-chemical-order roughened dislocation pathways},
	author       = {Li, Qing Jie and Sheng, Howard and Ma, Evan},
	journal      = {Nat Commun},
	volume       = 10,
	doi          = {10.1038/s41467-019-11464-7},
	date         = 2019,
	journal_full = {Nature Communications}
}

@article{lopanitsyna_modeling_2023,
	title        = {Modeling high-entropy transition metal alloys with alchemical compression},
	author       = {Lopanitsyna, Nataliya and Fraux, Guillaume and Springer, Maximilian A. and De, Sandip and Ceriotti, Michele},
	year         = 2023,
	journal      = {Phys. Rev. Materials},
	volume       = 7,
	doi          = {10.1103/PhysRevMaterials.7.045802},
	journal_full = {Physical Review Materials}
}

@misc{lovelyplots,
	title        = {{LovelyPlots, a collection of Matplotlib style sheets to nicely format figures for scientific papers, thesis and presentations}},
	author       = {Sheriff, Killian},
	year         = 2023,
	publisher    = {Zenodo},
	doi          = {10.5281/zenodo.10784160},
	url          = {https://github.com/killiansheriff/LovelyPlots},
	howpublished = {\url{https://github.com/killiansheriff/LovelyPlots}}
}

@article{MB_WC,
	title        = {Quantitative description of atomic architecture in solid solutions: a generalized theory for multicomponent short-range order},
	author       = {Ceguerra, Anna V. and Powles, Rebecca C. and Moody, Michael P. and Ringer, Simon P.},
	year         = 2010,
	journal      = {Phys. Rev. B},
	publisher    = {APS},
	volume       = 82,
	number       = 13,
	pages        = 132201,
	doi          = {10.1103/PhysRevB.82.132201},
	journal_full = {Physical Review B}
}

@article{metropolis_equation_2004,
	title        = {Equation of {State} {Calculations} by {Fast} {Computing} {Machines}},
	author       = {Metropolis, Nicholas and Rosenbluth, Arianna W. and Rosenbluth, Marshall N. and Teller, Augusta H. and Teller, Edward},
	journal      = {The Journal of Chemical Physics},
	volume       = 21,
	pages        = {1087--1092},
	doi          = {10.1063/1.1699114},
	date         = {2004-12},
	journal_full = {The Journal of Chemical Physics}
}

@misc{moniri_three-dimensional_2023,
      title={Three-dimensional atomic positions and local chemical order of medium- and high-entropy alloys}, 
      author={Saman Moniri and Yao Yang and Yakun Yuan and Jihan Zhou and Long Yang and Fan Zhu and Yuxuan Liao and Yonggang Yao and Liangbing Hu and Peter Ercius and Jun Ding and Jianwei Miao},
      year={2023},
      eprint={2305.14123},
      archivePrefix={arXiv},
      primaryClass={cond-mat.mtrl-sci},
      url={https://arxiv.org/abs/2305.14123}, 
}

@article{morris_weisfeiler_2019,
	title        = {Weisfeiler and {Leman} {Go} {Neural}: {Higher}-{Order} {Graph} {Neural} {Networks}},
	shorttitle   = {Weisfeiler and {Leman} {Go} {Neural}},
	author       = {Morris, Christopher and Ritzert, Martin and Fey, Matthias and Hamilton, William L. and Lenssen, Jan Eric and Rattan, Gaurav and Grohe, Martin},
	journal      = {AAAI},
	volume       = 33,
	pages        = {4602--4609},
	doi          = {10.1609/aaai.v33i01.33014602},
	date         = {2019-07},
	journal_full = {Proceedings of the AAAI Conference on Artificial Intelligence}
}

@book{munkres_elements_1993,
	title        = {Elements {Of} {Algebraic} {Topology}},
	author       = {Munkres, James R.},
	location     = {Boca Raton London New York},
	publisher    = {CRC Press},
	isbn         = {978-0-201-62728-2},
	date         = {1993-12},
	edition      = {First Edition}
}

@article{oses_high-entropy_2020,
	title        = {High-entropy ceramics},
	author       = {Oses, Corey and Toher, Cormac and Curtarolo, Stefano},
	year         = 2020,
	journal      = {Nat Rev Mater},
	volume       = 5,
	pages        = {295--309},
	doi          = {10.1038/s41578-019-0170-8},
	journal_full = {Nature Reviews Materials}
}

@article{owen_assessment_2017,
	title        = {An assessment of the lattice strain in the {CrMnFeCoNi} high-entropy alloy},
	author       = {Owen, L.R. and Pickering, E.J. and Playford, H.Y. and Stone, H.J. and Tucker, M.G. and Jones, N.G.},
	year         = 2017,
	journal      = {Acta Materialia},
	volume       = 122,
	pages        = {11--18},
	doi          = {10.1016/j.actamat.2016.09.032},
	journal_full = {Acta Materialia}
}

@article{owen_lattice_2018,
	title        = {Lattice distortions in high-entropy alloys},
	author       = {Owen, Lewis Robert and Jones, Nicholas Gwilym},
	year         = 2018,
	journal      = {J. Mater. Res.},
	volume       = 33,
	pages        = {2954--2969},
	doi          = {10.1557/jmr.2018.322},
	journal_full = {Journal of Materials Research}
}

@article{polak_local_2003,
	title        = {Local structures in medium-sized {Lennard}-{Jones} clusters: {Monte} {Carlo} simulations},
	shorttitle   = {Local structures in medium-sized {Lennard}-{Jones} clusters},
	author       = {Polak, Wieslaw and Patrykiejew, Andrzej},
	journal      = {Phys. Rev. B},
	volume       = 67,
	doi          = {10.1103/PhysRevB.67.115402},
	date         = {2003-03},
	journal_full = {Phys. Rev. B}
}

@article{polya_kombinatorische_1937,
	title        = {Kombinatorische {Anzahlbestimmungen} für {Gruppen}, {Graphen} und chemische {Verbindungen}},
	author       = {Pólya, G.},
	journal      = {Acta Math.},
	volume       = 68,
	pages        = {145--254},
	doi          = {10.1007/BF02546665},
	date         = {1937-01},
	journal_full = {Acta Mathematica}
}

@article{rackers_recipe_2023,
	title        = {A recipe for cracking the quantum scaling limit with machine learned electron densities},
	author       = {Rackers, Joshua A. and Tecot, Lucas and Geiger, Mario and Smidt, Tess E.},
	journal      = {Mach. Learn.: Sci. Technol.},
	volume       = 4,
	pages        = {015027},
	doi          = {10.1088/2632-2153/acb314},
	date         = {2023-02},
	journal_full = {Machine Learning: Science and Technology}
}

@article{rakita_mapping_2023,
	title        = {Mapping structural heterogeneity at the nanoscale with scanning nano-structure electron microscopy ({SNEM})},
	author       = {Rakita, Yevgeny and Hart, James L. and Das, Partha Pratim and Shahrezaei, Sina and Foley, Daniel L. and Mathaudhu, Suveen Nigel and Nicolopoulos, Stavros and Taheri, Mitra L. and Billinge, Simon J. L.},
	year         = 2023,
	journal      = {Acta Materialia},
	volume       = 242,
	pages        = 118426,
	doi          = {10.1016/j.actamat.2022.118426},
	journal_full = {Acta Materialia}
}

@incollection{robert_metropolishastings_2004,
	title        = {The {Metropolis}—{Hastings} {Algorithm}},
	author       = {Robert, Christian P. and Casella, George},
	booktitle    = {Monte {Carlo} {Statistical} {Methods}},
	location     = {New York, NY},
	publisher    = {Springer},
	series       = {Springer {Texts} in {Statistics}},
	doi          = {10.1007/978-1-4757-4145-2_7},
	isbn         = {978-1-4757-4145-2},
	date         = 2004,
	editor       = {Robert, Christian P. and Casella, George}
}

@article{shapeev_moment_2016,
	title        = {Moment {{Tensor Potentials}}: {{A Class}} of {{Systematically Improvable Interatomic Potentials}}},
	shorttitle   = {Moment {{Tensor Potentials}}},
	author       = {Shapeev, Alexander V.},
	journal      = {Multiscale Model. Simul.},
	volume       = 14,
	pages        = {1153--1173},
	doi          = {10.1137/15M1054183},
	date         = 2016,
	journal_full = {Multiscale Modeling \& Simulation},
	langid       = {english}
}

@article{sheriff2023quantifying,
  title={Quantifying chemical short-range order in metallic alloys},
  author={Sheriff, Killian and Cao, Yifan and Smidt, Tess and Freitas, Rodrigo},
  journal={Proceedings of the National Academy of Sciences},
  volume={121},
  number={25},
  pages={e2322962121},
  year={2024},
  publisher={National Acad Sciences}
}

@article{Singh2015,
	title        = {Atomic short-range order and incipient long-range order in high-entropy alloys},
	author       = {Singh, Prashant and Smirnov, A. V. and Johnson, D. D.},
	year         = 2015,
	journal      = {Phys. Rev. B},
	publisher    = {American Physical Society},
	volume       = 91,
	doi          = {10.1103/PhysRevB.91.224204},
	journal_full = {Phys. Rev. B}
}

@article{smidt_finding_2021,
	title        = {Finding symmetry breaking order parameters with {Euclidean} neural networks},
	author       = {Smidt, Tess E. and Geiger, Mario and Miller, Benjamin Kurt},
	journal      = {Phys. Rev. Research},
	volume       = 3,
	doi          = {10.1103/PhysRevResearch.3.L012002},
	date         = {2021-01},
	journal_full = {Physical Review Research}
}

@article{stukowski_visualization_2009,
	title        = {Visualization and analysis of atomistic simulation data with {OVITO}–the {Open} {Visualization} {Tool}},
	author       = {Stukowski, Alexander},
	journal      = {Modelling Simul. Mater. Sci. Eng.},
	volume       = 18,
	pages        = {015012},
	doi          = {10.1088/0965-0393/18/1/015012},
	date         = {2009-12},
	journal_full = {Modelling and Simulation in Materials Science and Engineering}
}

@article{sun_effect_2022,
	title        = {The effect of short-range order on mechanical properties of high entropy alloy {Al0}.{3CoCrFeNi}},
	author       = {Sun, Zerui and Shi, Changgen and Liu, Cuixia and Shi, Hang and Zhou, Jie},
	journal      = {Materials \& Design},
	volume       = 223,
	pages        = 111214,
	doi          = {10.1016/j.matdes.2022.111214},
	date         = {2022-11},
	journal_full = {Materials \& Design}
}

@article{utt_origin_2022,
	title        = {The origin of jerky dislocation motion in high-entropy alloys},
	author       = {Utt, Daniel and Lee, Subin and Xing, Yaolong and Jeong, Hyejin and Stukowski, Alexander and Oh, Sang Ho and Dehm, Gerhard and Albe, Karsten},
	journal      = {Nat Commun},
	volume       = 13,
	doi          = {10.1038/s41467-022-32134-1},
	date         = {2022-08},
	journal_full = {Nature Communications}
}

@article{varvenne_theory_2016,
	title        = {Theory of strengthening in fcc high entropy alloys},
	author       = {Céline Varvenne and Aitor Luque and William A. Curtin},
	year         = 2016,
	journal      = {Acta Materialia},
	volume       = 118,
	pages        = {164--176},
	doi          = {https://doi.org/10.1016/j.actamat.2016.07.040},
	journal_full = {Acta Materialia},
}

@article{walsh_extra_2023,
	title        = {Extra electron reflections in concentrated alloys do not necessitate short-range order},
	author       = {Walsh, Flynn and Zhang, Mingwei and Ritchie, Robert O. and Minor, Andrew M. and Asta, Mark},
	journal      = {Nat. Mater.},
	volume       = 22,
	pages        = {926--929},
	doi          = {10.1038/s41563-023-01570-9},
	date         = {2023-08},
	journal_full = {Nature Materials}
}

@article{walsh_magnetically_2021,
	title        = {Magnetically driven short-range order can explain anomalous measurements in {CrCoNi}},
	author       = {Walsh, Flynn and Asta, Mark and Ritchie, Robert O.},
	journal      = {Proc. Natl. Acad. Sci. U.S.A.},
	volume       = 118,
	doi          = {10.1073/pnas.2020540118},
	date         = {2021-03},
	journal_full = {Proceedings of the National Academy of Sciences}
}

@article{walsh_reconsidering_2023,
	title        = {Reconsidering short-range order in complex concentrated alloys},
	author       = {Walsh, Flynn and Abu-Odeh, Anas and Asta, Mark},
	year         = 2023,
	journal      = {MRS Bulletin},
	volume       = 48,
	pages        = {753--761},
	doi          = {10.1557/s43577-023-00555-y},
	journal_full = {MRS Bulletin}
}

@article{wang2023motif,
	title        = {Motif-Based Graph Representation Learning with Application to Chemical Molecules},
	author       = {Wang, Yifei and Chen, Shiyang and Chen, Guobin and Shurberg, Ethan and Liu, Hang and Hong, Pengyu},
	year         = 2023,
	journal      = {Informatics}
}

@article{wang_chemical_2022,
	title        = {Chemical medium-range order in a medium-entropy alloy},
	author       = {Wang, Jing and Jiang, Ping and Yuan, Fuping and Wu, Xiaolei},
	journal      = {Nat Commun},
	volume       = 13,
	doi          = {10.1038/s41467-022-28687-w},
	date         = {2022-02},
	journal_full = {Nature Communications}
}

@article{willatt_feature_2018,
	title        = {Feature optimization for atomistic machine learning yields a data-driven construction of the periodic table of the elements},
	author       = {Willatt, Michael J. and Musil, Félix and Ceriotti, Michele},
	year         = 2018,
	journal      = {Phys. Chem. Chem. Phys.},
	volume       = 20,
	pages        = {29661--29668},
	doi          = {10.1039/C8CP05921G},
	journal_full = {Physical Chemistry Chemical Physics}
}

@article{xie_percolation_2021,
	title        = {A percolation theory for designing corrosion-resistant alloys},
	author       = {Xie, Yusi and Artymowicz, Dorota M. and Lopes, Pietro P. and Aiello, Ashlee and Wang, Duo and Hart, James L. and Anber, Elaf and Taheri, Mitra L. and Zhuang, Houlong and Newman, Roger C. and Sieradzki, Karl},
	year         = 2021,
	journal      = {Nat. Mater.},
	volume       = 20,
	pages        = {789--793},
	doi          = {10.1038/s41563-021-00920-9},
	journal_full = {Nature Materials}
}

@misc{xu2019powerful,
	title        = {How Powerful are Graph Neural Networks?},
	author       = {Keyulu Xu and Weihua Hu and Jure Leskovec and Stefanie Jegelka},
	year         = 2019,
	archiveprefix = {arXiv},
	eprint       = {1810.00826},
	primaryclass = {cs.LG}
}

@article{xu_correlating_2023,
	title        = {Correlating local chemical and structural order using {Geographic} {Information} {Systems}-based spatial statistics},
	author       = {Xu, Michael and Kumar, Abinash and LeBeau, James M.},
	journal      = {Ultramicroscopy},
	volume       = 243,
	pages        = 113642,
	doi          = {10.1016/j.ultramic.2022.113642},
	date         = {2023-01},
	journal_full = {Ultramicroscopy}
}

@article{xu_determination_2023,
	title        = {Determination of short-range order in {TiVNbHf}({Al})},
	author       = {Xu, Michael and Wei, Shaolou and Tasan, C. Cem and LeBeau, James M.},
	journal      = {Applied Physics Letters},
	volume       = 122,
	doi          = {10.1063/5.0145289},
	date         = {2023-05},
	journal_full = {Applied Physics Letters}
}

@article{yang_coordination_2008,
	title        = {Coordination motifs and large-scale structural organization in atomic clusters},
	author       = {Yang, Zhu and Tang, Leihan},
	journal      = {Phys. Rev. B},
	volume       = 79,
	doi          = {10.1103/PhysRevB.79.045402},
	date         = {2008-12},
	journal_full = {Physical Review B}
}

@article{yin_atomistic_2021,
	title        = {Atomistic simulations of dislocation mobility in refractory high-entropy alloys and the effect of chemical short-range order},
	author       = {Yin, Sheng and Zuo, Yunxing and Abu-Odeh, Anas and Zheng, Hui and Li, Xiang-Guo and Ding, Jun and Ong, Shyue Ping and Asta, Mark and Ritchie, Robert O.},
	journal      = {Nat Commun},
	volume       = 12,
	doi          = {10.1038/s41467-021-25134-0},
	date         = {2021-08},
	journal_full = {Nature Communications}
}

@article{yin_neural_2021,
	title        = {Neural network-based order parameter for phase transitions and its applications in high-entropy alloys},
	author       = {Yin, Junqi and Pei, Zongrui and Gao, Michael C.},
	year         = 2021,
	journal      = {Nat Comput Sci},
	volume       = 1,
	pages        = {686--693},
	doi          = {10.1038/s43588-021-00139-3},
	journal_full = {Nature Computational Science}
}

@article{yu_theory_2022,
	title        = {Theory of history-dependent multi-layer generalized stacking fault energy— {A} modeling of the micro-substructure evolution kinetics in chemically ordered medium-entropy alloys},
	author       = {Yu, Peijun and Du, Jun-Ping and Shinzato, Shuhei and Meng, Fan-Shun and Ogata, Shigenobu},
	journal      = {Acta Materialia},
	volume       = 224,
	pages        = 117504,
	doi          = {10.1016/j.actamat.2021.117504},
	date         = {2022-02},
	journal_full = {Acta Materialia}
}

@article{zhang_short-range_2020,
	title        = {Short-range order and its impact on the {CrCoNi} medium-entropy alloy},
	author       = {Zhang, Ruopeng and Zhao, Shiteng and Ding, Jun and Chong, Yan and Jia, Tao and Ophus, Colin and Asta, Mark and Ritchie, Robert O. and Minor, Andrew M.},
	journal      = {Nature},
	volume       = 581,
	pages        = {283--287},
	doi          = {10.1038/s41586-020-2275-z},
	date         = {2020-05},
	journal_full = {Nature}
}

@article{zheng_multi-scale_2023,
	title        = {Multi-scale investigation of short-range order and dislocation glide in {MoNbTi} and {TaNbTi} multi-principal element alloys},
	author       = {Zheng, Hui and Fey, Lauren T. W. and Li, Xiang-Guo and Hu, Yong-Jie and Qi, Liang and Chen, Chi and Xu, Shuozhi and Beyerlein, Irene J. and Ong, Shyue Ping},
	journal      = {npj Comput Mater},
	volume       = 9,
	doi          = {10.1038/s41524-023-01046-z},
	date         = {2023-05},
	journal_full = {npj Computational Materials}
}

@article{zhou_atomic-scale_2022,
	title        = {Atomic-scale evidence of chemical short-range order in {CrCoNi} medium-entropy alloy},
	author       = {Zhou, Lingling and Wang, Qi and Wang, Jing and Chen, Xuefei and Jiang, Ping and Zhou, Hao and Yuan, Fuping and Wu, Xiaolei and Cheng, Zhiying and Ma, En},
	journal      = {Acta Materialia},
	volume       = 224,
	pages        = 117490,
	doi          = {10.1016/j.actamat.2021.117490},
	date         = {2022-02},
	journal_full = {Acta Materialia}
}

@software{repo_list,
	author       = {Sheriff, Killian and Cao, Yifan and Freitas, Rodrigo},
	url          = {https://github.com/stars/killiansheriff/lists/chemical-motif-characterization}
}

@software{cmi_github,
	author       = {Sheriff, Killian and Freitas, Rodrigo},
	url          = {https://github.com/killiansheriff/ChemicalMotifIdentifier}
}

@software{nshell_github,
	author       = {Sheriff, Killian and Freitas, Rodrigo},
	url          = {https://github.com/killiansheriff/NShellFinder}
}

@software{polya_github,
	author       = {Sheriff, Killian and Freitas, Rodrigo},
	url          = {https://github.com/killiansheriff/Polya}
}

@software{wc_github,
	author       = {Sheriff, Killian and Freitas, Rodrigo},
	url          = {https://github.com/killiansheriff/WarrenCowleyParameters}
}

@software{simplex_github,
	author       = {Sheriff, Killian and Freitas, Rodrigo},
	url          = {https://github.com/killiansheriff/simplex}
}

\end{document}